\documentclass[12pt, a4paper]{article}
\usepackage{mathrsfs}
\usepackage{bbm}
\usepackage{amsmath,amssymb}
\usepackage{color}
\usepackage{authblk}

\allowdisplaybreaks
\numberwithin{equation}{section}
\numberwithin{equation}{section} 

\newtheorem{thm}{Theorem}[section]
\newtheorem{prp}[thm]{Proposition}
\newtheorem{lem}[thm]{Lemma}
\newtheorem{defn}[thm]{Definition}
\newenvironment{dfn}{\begin{defn} \rm }{\end{defn}}
\newtheorem{cor}[thm]{Corollary}
\newtheorem{example}[thm]{Example}
\newenvironment{exa}{\begin{example} \rm }{ \end{example}}

\newtheorem{remark}[thm]{Remark}
\newenvironment{rmk}{\begin{remark} \rm }{\hfill $\Box$ \end{remark}}
\newenvironment{prf}{\noindent {\it Proof} \ }{\hfill $\Box$}
\newenvironment{prfof}[1]{\noindent {\it Proof of #1} \ }{\hfill $\Box$}

\newcommand{\eqa}{\begin{eqnarray}}
\newcommand{\eeqa}{\end{eqnarray}}
\newcommand{\beq}{\begin{equation}}
\newcommand{\eeq}{\end{equation}}

\newcommand\pd{\partial} \newcommand\od{\mathrm{d}}
\newcommand{\bm}[1]{\mbox{\boldmath{$#1$}}}
\newcommand{\nn}{\nonumber}

\newcommand{\ld}{\lambda} \newcommand{\Ld}{\Lambda}
\newcommand{\al}{\alpha}

\newcommand{\sg}{\sigma}
\newcommand{\om}{\omega} 
\newcommand{\Gm}{\Gamma}

\newcommand{\dt}{\delta}  
\newcommand{\ta}{\theta}

\newcommand{\vphi}{\varphi}

\newcommand{\diag}{\mathrm{diag}}

\newcommand\ad{\mathrm{ad}}
\newcommand{\res}{\mathrm{res}}

\newcommand\C{\mathbb{C}}

\newcommand\Z{\mathbb{Z}}
\newcommand\Zop{\mathbb{Z^{\mathrm{odd}}_+}}

\newcommand\cA{\mathcal{A}}
\newcommand\cD{\mathcal{D}}

\newcommand\cL{\mathcal{L}}

\newcommand\sL{\mathscr{L}}

\newcommand\fg{\mathfrak{g}}

\newcommand{\set}[1]{\left\{#1\right\}}
\newcommand{\bt}{\mathbf{t}}

\begin{document}
\title{Tau Function of the CKP Hierarchy and Non-linearizable Virasoro Symmetries}
\author[1]{Liang Chang}
\author[2]{ Chao-Zhong Wu}
\affil[1]{Department of Mathematics, UC Santa Barbara, CA 93106, USA
}

\affil[2]{Marie Curie fellow of the Istituto Nazionale di Alta
Matematica

SISSA, Via Bonomea 265, 34136 Trieste, Italy
}

\date{}
\maketitle

\begin{abstract}
We introduce a single tau function that represents the C-type
Kadomtsev-Petviashvili (CKP) hierarchy into a generalized Hirota
``bilinear'' equation. The actions on the tau function by additional
symmetries for the hierarchy are also calculated, which involve
strictly more than a central extension of the $w^C_\infty$-algebra.
As an application, for Drinfeld-Sokolov hierarchies associated to
affine Kac-Moody algebras of type C, we obtain a formula to compute
the obstacles in linearizing their Virasoro symmetries and hence
prove the Virasoro symmetries to be non-linearizable when acting on
the tau function.

\vskip 2ex \noindent{\bf Key words}: tau function; CKP hierarchy;
Drinfeld--Sokolov hierarchy; Virasoro symmetry
\end{abstract}

\section{Introduction}

The Kadomtsev-Petviashvili (KP) hierarchy together with its
subhierarchies of types B and C \cite{DKJM-KPBKP, DJKM-CKP},
abbreviated as the BKP and the CKP hierarchies respectively, has
attracted much research interest in areas of mathematical physics.
These hierarchies can be represented equivalently as Lax equations
of pseudo-differential operators or as bilinear equations. For
instance, the CKP hierarchy concerned in the present paper is
defined by the following bilinear equation
\begin{equation}\label{ckp0}
\res_{z}  w(\mathbf{t};z)w(\mathbf{t}';-z)=0,
\end{equation}
where $w$ is the so-called wave function depending on the time
variable $\mathbf{t}=(t_1,t_3,t_5,\dots)$ and a parameter $z$, and
$\res_z\sum_{i}f_i z^i=f_{-1}$ for any formal Laurent series in $z$.

For the CKP hierarchy, in contrast to the KP and the BKP cases, it
seems not to exist a single tau function that represents
\eqref{ckp0} to the form of Hirota bilinear equations, though it was
pointed out by Date, Jimbo, Kashiwara and Miwa \cite{DJKM-CKP} that
a tau function may be constructed from the action of bosonic fields
on the vacuum vector in a Fock space. The idea in \cite{DJKM-CKP}
was developed by van de Leur, Orlov and Shiota \cite{vdLOS} later.
They introduced a series of fermionic operators besides the bosonic
fields, and constructed a tau function depending on both time
variables $t_k$ and certain odd Grassmannian parameters. In fact it
is a tau function of a generalization of the CKP hierarchy, i.e., a
system of bilinear equations like \eqref{ckp0} of wave functions
labeled with ``odd number of Odd Partitions with Distinct parts''
(see equation~(2.38) in \cite{vdLOS}). In particular, their wave
function $w_1$ (with its expansion (2.35)--(2.37) in \cite{vdLOS})
solves the bilinear equation \eqref{ckp0} of the CKP hierarchy, but
it is related to a series of tau functions rather than only one.

In this paper we are to introduce a tau function $\tau(\bt)$ of the
CKP hierarchy by making use of its Hamiltonian densities, in
consideration of that the hierarchy carries a series of
bi-Hamiltonian structures reduced from those for the KP hierarchy
\cite{Dickey}. This tau function will be shown related to the wave
function via the following formula
\begin{equation}\label{wtau0}
w(\bt;z)=\left(1+\frac{1}{z}\frac{\pd}{\pd
t_1}\log\frac{G(z)\tau(\bt)}{\tau(\bt)}\right)^{1/2}
\frac{G(z)\tau(\bt)}{\tau(\bt)}e^{\xi(\bt;z)},
\end{equation}
where
\[
G(z)=\exp\left(-\sum_{k\in\Zop}\frac{2}{k\,z^k}\frac{\pd}{\pd
t_k}\right), \quad \xi(\mathbf{t}; z)=\sum_{k\in\Zop} t_k z^k.
\]
Observe that \eqref{wtau0} is different from those formulae relating
wave and tau functions of integrable hierarchies in the literature,
namely, now there is a square-root factor depending on the tau
function. By substituting this formula into \eqref{ckp0}, the CKP
hierarchy is recast to a generalized Hirota ``bilinear'' equation
(see equation~\eqref{taubl} below).

As the tau function of the CKP hierarchy is introduced, we continue
to study the action on it by the additional symmetries.  Recall that
the additional symmetries for the CKP hierarchy were constructed by
He, Tian, Foerster and Ma \cite{HTFM} with the help of certain
Orlov-Schulman operators \cite{OS}. They also showed that these
additional symmetries acting on the wave function $w(\bt;z)$ form a
centerless $w^C_\infty$-algebra. However, their actions on tau
function of the hierarchy still need to be clarified. As observed by
Adler, Shiota and van Moerbeke \cite{ASvM} in the context of matrix
integrals, lifting the actions of additional symmetries on wave
function to that on tau function results in a central extension of
the sort of $w_\infty$-algebra. Such phenomenon were confirmed for
the KP and the BKP hierarchies as well as for the two-dimensional
Toda lattice and the two-component BKP hierarchies \cite{ASvM, Di,
vdL, Tu, Wu}. In contrast to them, a counter example will be found,
that is the CKP hierarchy \eqref{ckp0}. More exactly, for the CKP
hierarchy when the additional symmetries act on its tau function
given in \eqref{wtau0}, it implies not only a central extension of
the $w^C_\infty$-algebra but also some non-trivial ``tails'' given
by polynomials in at-least-second-order derivatives of $\log\tau$
with respective to the time variables. So far as we know, such kind
of counter examples have not been considered in the literature
before, and whether their property can be illustrated in matrix
models is unclear.

This paper is also motivated by the study of non-linearizable
Virasoro symmetries for integrable hierarchies proposed recently by
one of the authors in \cite{Wu-DS} when considering Drinfeld-Sokolov
hierarchies. Recall that for every affine Kac-Moody algebra with an
arbitrary vertex of the Dynkin diagram marked (only the case of the
zeroth vertex is considered below), Drinfeld and Sokolov \cite{DS}
constructed an integrable hierarchy of Korteweg-de Vries (KdV) type.
These hierarchies are applied to various areas of mathematical
physics \cite{DZ, Kon, dV, Wi}. For instance, the Drinfeld-Sokolov
hierarchy associated to each simply-laced affine Kac-Moody algebra
were conjectured by Dubrovin and Zhang \cite{DZ} to coincide with
the topological integrable hierarchies constructed from semisimple
Frobenius manifolds corresponding to ADE-type Coxeter groups. By now
this conjecture has been proved in \cite{DLZ} and \cite{Wu-DS}, see
also \cite{DZ, LWZ, Wu}. One of the main steps in the proof is to
check the linearization of Virasoro symmetries acting on the tau
function. In fact, in \cite{Wu-DS} we obtained a general formula to
compute how the tau function is acted by Virasoro symmetries for all
Drinfeld-Sokolov hierarchies, and confirmed the linearization
property for each hierarchy associated to an either simply-laced  or
twisted \cite{Kac} affine Kac-Moody algebra.
 In contrast, Drinfeld-Sokolov hierarchies associated to
affine Kac-Moody algebras of the other types were claimed
\cite{Wu-DS} not to have linearized Virasoro symmetries. We will
prove this claim for the hierarchies of type $C_n^{(1)}$, that is
\begin{thm} \label{thm-dsc}
For the Drinfeld-Sokolov hierarchy associated to affine Kac-Moody
algebra of type $C_n^{(1)}$ with $n\ge2$, the Virasoro symmetries
acting on the tau function (see \eqref{tauvir} below) are
non-linearizable. More precisely, the obstacles in linearizing the
Virasoro symmetries, which were introduced in \cite{Wu-DS}, can be
computed via formula \eqref{OT} below.
\end{thm}
This theorem shows that the Drinfeld-Sokolov hierarchy of type
$C_n^{(1)}$ with $n\ge2$ does not belong to the set of topological
hierarchies associated to semisimple Frobenius manifolds \cite{DZ}.
The proof of it is based on the observation that each of such
hierarchies is equivalent to a certain reduction of the CKP
hierarchy and that its Virasoro symmetries can be reduced
accordingly from the additional symmetries for the latter.

To achieve the above results, we will arrange this paper as follows.
In next section we recall the definition of the CKP hierarchy and
its additional symmetries. In Section~3 we introduce a tau function
of the CKP hierarchy by using its Hamiltonian densities and then
represent the hierarchy into a ``bilinear'' equation of tau
function. The actions on tau function by the additional symmetries
are considered in Section~4. Section~5 consists of two parts. The
first part is devoted to a brief review of the construction of the
Drinfeld-Sokolov hierarchies from affine Kac-Moody algebras of type
C as well as their tau function and Virasoro symmetries; in the
second part, these hierarchies and their Virasoro symmetries are
reconstructed from a $2n$-reduction of the CKP hierarchy and its
additional symmetries, which provides an alternative way to compute
the obstacles in linearizing the Virasoro symmetries considered in
\cite{Wu-DS}. A summary will be given in the final section.

\section{The CKP hierarchy and its additional symmetries}

Let $\cA$ be an algebra of smooth functions of a spatial coordinate
$x$, and $D=\od/\od x$ be a derivation on $\cA$. The algebra of
pseudo-differential operators is the following linear space
\begin{equation}\label{cD}
\cD=\set{\sum_{i<\infty} f_i D^i\mid f_i\in\cA}
\end{equation}
equipped with a product defined by
\[
f D^i\cdot g D^j=\sum_{r\geq0}\binom{i}{r}f\, D^r(g)\, D^{i+j-r},
\quad f,g\in\cA.
\]
For any operator $A=\sum_{i} f_i D^i\in\cD$, its nonnegative part,
negative part, residue and adjoint operator are given respectively
by
\begin{align}\label{Apm}
&A_+=\sum_{i\geq0} f_i  D^i, \quad A_-=\sum_{i<0} f_i  D^i, \quad
\res\,A=f_{-1},\quad A^*=\sum_{i}(- D)^i f_i.
\end{align}
These notions will be frequently used in the sequel.

Assume a pseudo-differential operator
\begin{equation} \label{Lckp}
L= D+\sum_{i\ge1}v_i D^{-i}\in\cD
\end{equation}
satisfies $L^*=-L$. Note that each coefficient $v_{2 j}$ is a
differential polynomial in the functions $v_1$, $v_3$, \dots, $v_{2
j-1}$. The CKP hierarchy is defined by the following Lax equations:
\begin{equation}\label{ckp}
 \frac{\pd L}{\pd t_k}=[(L^k)_+, L],\quad k\in\Zop,
\end{equation}
which form a system of evolutionary equations of the vector function
$\mathbf{v}=(v_1, v_3, v_5, \dots)$ depending on
$\bt=(t_1,t_3,t_5,\dots)$. Clearly $\pd/\pd t_1=\pd/\pd x$;
henceforth we simply assume $t_1=x$.

The operator $L$ can be represented in a dressing form as
\begin{equation} \label{Ldr}
L=\Phi D\Phi^{-1},
\end{equation}
where $\Phi$ is a pseudo-differential operator:
\begin{equation} \label{Phi}
\Phi=1+\sum_{i\ge 1}a_i D^{-i}, \quad \Phi^*=\Phi^{-1}.
\end{equation}
Note that the dressing operator $\Phi$ is determined up to
multiplication to the right by an arbitrary operator of the form
\eqref{Phi} with constant coefficients. With the help of the
dressing operator, the CKP hierarchy \eqref{ckp} can be redefined by
the Sato equations:
\begin{equation}\label{Phit}
\frac{\pd \Phi}{\pd t_k}=- (L^k)_-\Phi, \quad k\in\Zop.
\end{equation}

Let $\xi(\mathbf{t}; z)=\sum_{k\in\Zop} t_k z^k$ with some parameter
$z$. Introduce a wave function
\begin{align}\label{wavef}
w(\mathbf{t}; z)=\Phi
e^{\xi(\mathbf{t};z)}=\phi(\bt;z)e^{\xi(\mathbf{t};z)},
\end{align}
where
\begin{equation} \label{phiz}
\phi(\bt;z)=1+\sum_{i\ge 1}a_i z^{-i}
\end{equation}
(the convention $D^i e^{x z}=z^i e^{x z}$ for any integer $i$ is
adopted). The dual wave function reads
\begin{equation}\label{}
w^*(\mathbf{t}; z)=(\Phi^{-1})^*
e^{-\xi(\mathbf{t};z)}=\phi(\bt;-z)e^{-\xi(\mathbf{t};z)}=w(\mathbf{t};
-z).
\end{equation}
The CKP hierarchy \eqref{Phit}, or \eqref{ckp}, is equivalent to the
following bilinear equation \cite{DJKM-CKP}:
\begin{equation}\label{blckp}
\res_{z}  w(\mathbf{t};z)w(\mathbf{t}';-z)=0.
\end{equation}
Here $\res_z\sum_{i}f_i z^i=f_{-1}$ for any formal Laurent series
$\sum_{i}f_i z^i$ in $z$.

For the CKP hierarchy, He, Tian, Foerster and Ma \cite{HTFM}
constructed its additional symmetries by using the following
Orlov-Schulman \cite{OS} operator:
\[
M=\Phi\Gm\Phi^{-1}, \quad \Gm=\sum_{k\in\Zop}k\,t_k  D^{k-1}.
\]
Clearly $[L, M]=1$.

\begin{rmk}
Strictly speaking, the operator $M$ does not belong to the algebra
$\cD$ in \eqref{cD} for $M$ may contain infinitely many terms with
positive power in $D$. A trial to resolve this problem was given in
\cite{Wu}, that is to assign certain degrees to $t_k$ and extend
$\cD$ to be the so-called algebra of pseudo-differential operators
of the first type (cf. \cite{LWZ}). In this way $L$ and $M$ are
contained in a common algebra so that the product between them makes
sense.
\end{rmk}

Given any pair of integers $(m,l)$ with $m\ge0$, let
\begin{equation}\label{}
 A_{m l}=M^m
L^l-(-1)^l L^{l}M^m.
\end{equation}
In particular, one can check
\begin{align*}
A_{0 l}=&\left\{\begin{array}{cl}
                 0,  & l \hbox{ even; } \\
                 2 L^l, \quad & l \hbox{ odd },
               \end{array}
               \right. \\
A_{1 l}=&\left\{\begin{array}{cl}
                 -l\,L^{l-1}, & l \hbox{ even; } \\
                 2 M\,L^l+l\,L^{l-1}, \quad & l \hbox{ odd }.
               \end{array}
               \right.
\end{align*}
Note also $A_{m l}^*=- A_{m l}$, hence there are constants $c_{m l,
m' l'}^{q r}$ such that
\begin{equation}\label{}
[A_{m l}, A_{m' l'}]=\sum_{q,r}c_{m l, m' l'}^{q r}A_{q r}.
\end{equation}
In other words, all operators $A_{m l}$ generate a centerless
$w^C_{\infty}$-algebra. As a matter of fact, only those  $A_{m l}$
with odd indices $l$ are linearly independent, and the above
structure constants are uniquely determined by letting $c_{m l, m'
l'}^{q r}=0$ for even $r$. For example, one has
\begin{align*}
&c_{0l,0l'}^{q r}=c_{0,l;1,2i}^{q r}=0, \quad
c_{0,2 i+1;1,2 j+1}^{q r}=2(2 i+1)\dt_{q0}\dt_{r,2(i+j)+1}, \\
&c_{1,2 i+1;1,2 j+1}^{q r}=4(i-j)\dt_{q1}\dt_{r,2(i+j)+1}.
\end{align*}

The following equations are well defined:
\begin{equation} \label{sml}
\frac{\pd \Phi}{\pd s_{m l}}=- (A_{m l})_-\Phi, \quad m\ge0,
~l\in\Z.
\end{equation}
These flows are assumed to commute with $\pd/\pd x$.
\begin{prp}[\cite{HTFM}]\label{thm-st}
The flows \eqref{sml} commute with those in \eqref{Phit} that
compose the CKP hierarchy. Moreover, the vector fields $\pd/\pd s_{m
l}$ acting on the dressing operators $\Phi$ (or on the wave function
$w(\bt;z)$) satisfy
\begin{equation}\label{}
\left[\frac{\pd}{\pd s_{m l}}, \frac{\pd}{\pd s_{m'
l'}}\right]=-\sum_{q,r}c_{m l, m' l'}^{q r}\frac{\pd}{\pd s_{q r}}.
\end{equation}
\end{prp}
This proposition means that, equations \eqref{sml} define a set of
symmetries, named as additional symmetries, for the CKP hierarchy.
These additional symmetries acting on the wave function form a
centerless $w^C_{\infty}$-algebra.

Introduce a generating function of operators as
\begin{equation}\label{Y}
Y(\ld,
\mu)=\sum_{m=0}^\infty\frac{(\mu-\ld)^m}{m!}\sum_{l=-\infty}^\infty
\ld^{-m-l-1}(A_{m,m+l})_-
\end{equation}
with parameters $\ld$ and $\mu$. He et al \cite{HTFM} obtained the
following (cf. the case of the KP hierarchy studied in \cite{Or})
\begin{prp}\label{thm-Y}
The generator \eqref{Y} can be represented as
\begin{equation}\label{Yw}
Y(\ld,\mu)=w(\bt;-\ld)D^{-1}w(\bt;\mu)+w(\bt;\mu)D^{-1}w(\bt;-\ld).
\end{equation}
\end{prp}

\section{Tau function of the CKP hierarchy}

We are to introduce a tau function of the CKP hierarchy. To this end
let us first rewrite the hierarchy \eqref{ckp} into the form of
Hamiltonian systems.

Given an arbitrary positive integer $n$, the operator $L$ in
\eqref{Lckp} satisfies $(L^{2n})^*=L^{2n}$. Assume $F$ to be a
formal functional depending on $L$:
\begin{equation}\label{F}
F=\int f\left(\mathbf{v}, \pd_x\mathbf{v}, \pd_x^2\mathbf{v},
\dots\right)\,\od x \in \cA/\pd_x\cA.
\end{equation}
Its variational derivative with respect to $L^{2n}$ is defined to be
a pseudo-differential operator $P$ such that
\[
\dt F=\int\res (P\,\dt L^{2n})\,\od x, \quad P^*=-P.
\]
Let $F_P$ denote the functional whose variational derivative with
respect to $L^{2n}$ is $P\in\cD$. For such functionals, there is a
pair of compatible Poisson brackets that are reduced from those in
the bi-Hamiltonian representations of the KP hierarchy (see
\cite{Dickey} and references therein):
\begin{align}\label{pb1}
\set{F_P, F_Q}_1^n=&\int\res\left(P([-Q_-, L^{2n}]+[Q,
L^{2n}]_-)\right)\od x, \\
\set{F_P, F_Q}_2^n=&\int\res\left(P(-(L^{2n}Q)_-
L^{2n}+L^{2n}(Q\,L^{2n})_-)\right)\od x. \label{pb2}
\end{align}
Then the CKP hierarchy \eqref{ckp} can be represented in a
bi-Hamiltonian recursive form as
\begin{equation}\label{biHamckp}
\frac{\pd F}{\pd t_k}=\set{F, H_{k+2n}}_1^n=\set{F, H_{k}}_2^n,
\quad k\in\Zop,
\end{equation}
where $F$ is an arbitrary functional of the form \eqref{F}, and the
Hamiltonians are
\begin{equation}\label{Hk}
H_k=\frac{2\,n}{k}\int\res\,L^k\,\od x.
\end{equation}

The Hamiltonian densities in \eqref{Hk} are tau-symmetric \cite{DZ}.
That is to say, they define a closed $1$-form
\[
\om=\sum_{k\in\Zop} \res\,L^k\,\od t_k,
\]
hence there locally is a smooth function $\tau(\bt)$ such that
\begin{equation}\label{tau}
\od\left(2\,\pd_x\,\log\tau\right)=\om.
\end{equation}
More precisely, we have
\begin{dfn}
Given any solution of the CKP hierarchy \eqref{ckp}, by tau function
we mean a smooth function $\tau(\bt)$ that satisfies
\begin{equation}\label{tauckp}
\frac{\pd^2\log\tau}{\pd t_k\,\pd
t_l}=\frac{1}{2}\pd_x^{-1}\res[(L^k)_+, L^l], \quad k, l\in\Zop.
\end{equation}
Here on the right hand side of \eqref{tauckp} the integration
constants are taken to be zero (the residue of any commutator of
pseudo-differential operators is a total derivative in $x$).
\end{dfn}
Observe that $\log\tau$ is determined up to addition of a linear
function of the time variables.

In order to relate the tau function to the wave function of the CKP
hierarchy, we introduce the following shift operator
\[
G(\bt;z)=\exp\left(-\sum_{k\in\Zop}\frac{2}{k\,z^k}\frac{\pd}{\pd
t_k}\right).
\]
We also write $G(z)=G(\bt;z)$ in case no confusion would happen.
\begin{prp}\label{prp-wtau}
For the CKP hierarchy, the tau function in \eqref{tauckp} and the
wave function \eqref{wavef} are related via the following formula
\begin{equation}\label{wtau}
w(\bt;z)=\left(1+\frac{1}{z}\pd_x\log\frac{G(z)\tau(\bt)}{\tau(\bt)}\right)^{1/2}
\frac{G(z)\tau(\bt)}{\tau(\bt)}e^{\xi(\bt;z)}.
\end{equation}
\end{prp}

To prove this proposition, we need the following two lemmas.
\begin{lem}
Recall the dressing operator \eqref{Phi} for the CKP hierarchy. It
holds that
\begin{equation}\label{a1tau}
a_1(\bt)=-2\,\pd_x\log\tau(\bt).
\end{equation}
\end{lem}
\begin{prf}
One takes the residue of equation \eqref{Phit}, then the lemma
follows from the definition of the tau function in \eqref{tau}.
\end{prf}

\begin{lem}
Let $\vphi(\bt;\ld)=\phi(\bt;\ld)G(\ld)\phi(\bt;-\ld)$ (recall
$\phi$ in \eqref{phiz}) with $\ld$ being a parameter. Then it
satisfies
\begin{itemize}
\item[(i)]
\begin{equation}\label{fa1}
\vphi(\bt;\ld)=1+\frac{1}{2\ld}(1-G(\ld))a_1(\bt);
\end{equation}
\item[(ii)]
\begin{equation}\label{fa2}
2\,\pd_x\log\phi(\bt;\ld)-\pd_x\log\,\vphi(\bt;\ld)
=(1-G(\ld))a_1(\bt).
\end{equation}
\end{itemize}
\end{lem}
\begin{prf}
According to the bilinear equation \eqref{blckp}, we have
\begin{align}
0=&\res_z \phi(\bt;z)e^{\xi(\bt;z)}G(\ld)\left(\phi(\bt;-z)e^{-\xi(\bt;z)}\right) \nn\\
=&\res_z \phi(\bt;z)G(\ld)\phi(\bt;-z)\frac{1+z/\ld}{1-z/\ld} \nn\\
=&\ld\left.\left(\phi(\bt;z)G(\ld)\phi(\bt;-z)
    \left(1+\frac{z}{\ld}\right)\right)_{-}\right|_{z=\ld} \nn\\
=&\ld\left(
\phi(\bt;z)G(\ld)\phi(\bt;-z)\left(1+\frac{z}{\ld}\right)
    -1-(1+a_1(\bt)z^{-1}-G(\ld)a_1(\bt)z^{-1})\frac{z}{\ld}\right)_{z=\ld} \nn\\
=&2\ld\phi(\bt;\ld)G(\ld)\phi(\bt;-\ld)-2\ld-(1-G(\ld))a_1(\bt),
\end{align}
in the third equality of which the subscript ``$-$'' means to take
the negative-power part of a series in $z$. Thus the first formula
\eqref{fa1} is valid.

Secondly, the bilinear equation \eqref{blckp} also yields
\begin{align}
0=&\res_z\left(\pd_x(\phi(\bt;z)e^{\xi(\bt;z)})\right)
    G(\ld)\left(\phi(\bt;-z)e^{-\xi(\bt;z)}\right) \nn\\
=&\res_z (z\phi(\bt;z)+\pd_x\phi(\bt;z))G(\ld)\phi(\bt;-z)\frac{1+z/\ld}{1-z/\ld} \nn\\
=&\ld\left.\left((z\phi(\bt;z)+\pd_x\phi(\bt;z))G(\ld)\phi(\bt;-z)
    \left(1+\frac{z}{\ld}\right)\right)_{-}\right|_{z=\ld} \nn\\
=&2\,\ld(\ld\phi(\bt;\ld)+\pd_x\phi(\bt;\ld))G(\ld)\phi(\bt;-\ld)
\nn\\
& -\ld\cdot\ld(1+a_1(\bt)\ld^{-1}-G(\ld)a_1(\bt)\ld^{-1}) \nn
\\
&-\ld\cdot\ld\big(1+a_1(\bt)\ld^{-1}+a_2(\bt)\ld^{-2}-G(\ld)a_1(\bt)\ld^{-1}
+G(\ld)a_2(\bt)\ld^{-2}
\nn \\
&-a_1(\bt)G(\ld)a_1(\bt)\ld^{-2}\big) -\ld \pd_x a_1(\bt)\ld^{-1}
\nn\\
=& 2\,\ld^2 \vphi(\bt;\ld)+2\ld \pd_x\phi(\bt;\ld)\cdot
G(\ld)\phi(\bt;-\ld)
-2\ld^2 \nn\\
&-2\ld(1-G(\ld))a_1(\bt)-(1+G(\ld))a_2(\bt)+a_1(\bt)G(\ld)a_1(\bt)
-\pd_x a_1(\bt). \label{eq2}
\end{align}
On the other hand, it follows from \eqref{Phi} that
$2\,a_2(\bt)=a_1(\bt)^2-\pd_x a_1(\bt)$, hence
\begin{align}
&2(1+G(\ld)) a_2(\bt) \nn\\
=&\left((1-G(\ld)) a_1(\bt)\right)^2+2\,a_1(\bt)G(\ld)
a_1(\bt)-(1+G(\ld))\pd_x a_1(\bt).\label{a1a2}
\end{align}
Substituting \eqref{fa1} and \eqref{a1a2} into \eqref{eq2}, one
deduces
\begin{align}
&2\ld\pd_x\phi(\bt;\ld)\cdot G(\ld)\phi(\bt;-\ld) \nn\\
=&\ld(1-G(\ld))a_1(\bt)+\frac{1}{2}\left((1-G(\ld))a_1(\bt)\right)^2
+\frac{1}{2}(1-G(\ld))\pd_x a_1(\bt) \nn\\
=& \ld\,\vphi(\bt;\ld)(1-G(\ld))a_1(\bt)+\ld\pd_x \vphi(\bt;\ld).
\end{align}
Divide both sides by $\ld\,\vphi(\bt;\ld)$, then we obtain
\eqref{fa2}. The lemma is proved.
\end{prf}
\\

\begin{prfof}{Proposition~\ref{prp-wtau}}
Substituting \eqref{a1tau} into \eqref{fa1} and \eqref{fa2}, we have
respectively
\begin{align}\label{ftau}
& \vphi(\bt;\ld)=1+\frac{1}{\ld}(G(\ld)-1)\pd_x\log\tau(\bt)
=1+\frac{1}{\ld}\pd_x\log\frac{G(\ld)\tau(\bt)}{\tau(\bt)}, \\
&
\phi(\bt;\ld)=\sqrt{\vphi(\bt;\ld)}\,\frac{G(\ld)\tau(\bt)}{\tau(\bt)}.
\label{phitau}
\end{align}
They, with $\ld$ replaced by $z$, lead to the equality \eqref{wtau}
by virtue of the definition of the wave function \eqref{wavef}.
Therefore the proposition is proved.
\end{prfof}

\begin{rmk}
In \cite{vdLOS}, van de Leur et al found a formula that represents
the wave function of the CKP hierarchy to some fermionic tau
functions. This formula reads (see equations (2.25), (2.35) and
(2.37) in \cite{vdLOS})
\begin{equation}\label{wtauLOS}
w(\bt;z)=\frac{G(-z)\left(\tau_0(\bt)+ \sum_{i\ge2}\tau_{(2
i-1,1)}(\bt)z^{-i}\right) }{\tau_0(\bt)}e^{\xi(\bt;z)}.
\end{equation}
On the other hand, our formula \eqref{wtau} can be rewritten as
\begin{equation}\label{wtau2}
w(\bt;z)=
\frac{G(z)\left(\tau(\bt)\sqrt{(1+{z}^{-1}(1-G(-z))\pd_x\log\tau(\bt)}\right)
}{\tau(\bt)}e^{\xi(\bt;z)}.
\end{equation}
Note that $(1-G(-z))\pd_x\log\tau(\bt)=O(1/z)$. By comparting the
coefficients of $w(\bt;z)e^{-\xi(\bt;z)}$ given in \eqref{wtauLOS}
and \eqref{wtau2}, we obtain
\begin{equation}\label{}
\pd_x\log\tau_0(\bt)=-\pd_x\log\tau(\bt),
\end{equation}
hence $\tau_0(\bt)\tau(\bt)=\mathrm{const}$; accordingly, $\tau_{(2
i-1,1)}(\bt)$ with $i\ge2$ can be recursively represented by
$\tau(\bt)$ and its derivatives. Thus the construction in
\cite{vdLOS} (see also \cite{DJKM-CKP}) in fact provides a fermionic
interpretation for the tau function $\tau(\bt)$ defined via
Hamiltonian densities.
\end{rmk}

Denote
\begin{equation}\label{Xz}
X(\bt;z)=e^{\xi(\bt;z)}G(\bt;z).
\end{equation}
Now we achieve the main result of the present section by
substituting \eqref{wtau} into \eqref{blckp}.
\begin{thm}
The CKP hierarchy \eqref{ckp} is equivalent to the following
``bilinear'' equation of tau function:
\begin{equation}\label{taubl}
\res_z
\sqrt{\vphi(\bt;z)\vphi(\bt';-z)}\,X(\bt;z)\tau(\bt)X(\bt';z)\tau(\bt')=0,
\end{equation}
where the function $\vphi(\bt;z)$ is given in \eqref{ftau}.
\end{thm}

Observe the difference between \eqref{taubl} and those Hirota
bilinear equations of the usual sense (for example, the KP and the
BKP hierarchies \cite{DKJM-KPBKP}) in the literature: now there is a
square-root factor given by the tau function!

At the end of this section, we note that $\vphi(\bt;z)=1+O(1/z^2)$
as $z\to\infty$, and that it satisfies
\begin{equation}\label{Gvphi}
G(-z)\vphi(\bt;z)=\vphi(\bt;-z).
\end{equation}
These properties will be employed in the forthcoming section.

\section{Additional symmetries represented via tau function}

In this section we want to represent the additional symmetries
\eqref{sml} for the CKP hierarchy via the tau function introduced
above. Our main tool is served by vertex operators.

For any positive odd integer $k$, denote
\[
p_k=2\frac{\pd}{\pd t_k}, \quad p_{-k}=k\,t_k.
\]
Clearly $[p_k,p_l]=2\,k\,\dt_{k,-l}$ with $k, l\in\Zop$. Introduce a
vertex operator
\begin{equation}\label{X}
X(\bt;\ld,\mu)=\,:\exp\left(\sum_{k\in
\Z^{\mathrm{odd}}}\frac{p_k}{k\,\ld^k}-\sum_{k\in
\Z^{\mathrm{odd}}}\frac{p_k}{k\,\mu^k}\right):\,,
\end{equation}
where the normal-order product ``$:\ :$'' means to place $p_{k>0}$
to the right of $p_{k<0}$, and $\ld$ and $\mu$ are parameters.
Without any confusion we will simply write
$X(\ld,\mu)=X(\bt;\ld,\mu)$.

\begin{lem}\label{thm-Xtau}
The tau function in \eqref{tauckp} of the CKP hierarchy satisfies
the following equality
\begin{align}\label{Xtau}
&\pd_x\frac{X(\ld,\mu)\tau(\bt)}{\tau(\bt)} \nn\\
=&\frac{\mu-\ld}{\mu+\ld}w(\bt;\mu)w(\bt;-\ld) \left(\ld
G(\mu)\sqrt{\frac{\vphi(\bt;-\ld)}{\vphi(\bt;-\mu)}}+\mu
G(-\ld)\sqrt{\frac{\vphi(\bt;\mu)}{\vphi(\bt;\ld)}}\right),
\end{align}
where the functions $w(\bt;z)$ and $\vphi(\bt;z)$ are given by
\eqref{wtau} and \eqref{ftau} respectively.
\end{lem}
\begin{prf}
We write
\[
X(\ld,\mu)=X(\bt;\ld,\mu)=e^{-\xi(\bt;\ld)+\xi(\bt;\mu)}G(-\ld)G(\mu).
\]
Recalling \eqref{Xz}, it is straightforward to verify
\begin{align*}
&X(\bt;z)X(\bt;\ld,\mu)=\frac{1-{\mu}/{z}}{1+{\mu}/{z}}\frac{1+{\ld}/{z}}{1-{\ld}/{z}}
e^{\xi(\bt;z)-\xi(\bt;\ld)+\xi(\bt;\mu)}G(z)G(-\ld)G(\mu),
\\
&X(\bt;\ld,\mu)X(\bt;z)=\frac{1-{z}/{\mu}}{1+{z}/{\mu}}\frac{1+{z}/{\ld}}{1-{z}/{\ld}}
e^{\xi(\bt;z)-\xi(\bt;\ld)+\xi(\bt;\mu)}G(z)G(-\ld)G(\mu).
\end{align*}
The bilinear equation \eqref{taubl} yields
\begin{align}\label{}
0=&\res_z
X(\bt;\ld,\mu)\sqrt{\vphi(\bt;z)\vphi(\bt';-z)}X(\bt;z)\tau(\bt)X(\bt';-z)\tau(\bt')
\nn
\\
=&\res_z\sqrt{G(-\ld)G(\mu)\vphi(\bt;z)\cdot \vphi(\bt';-z)}
X(\bt;\ld,\mu)X(\bt;z)\tau(\bt)X(\bt';-z)\tau(\bt') \nn
\\
=&\res_z\sqrt{G(-\ld)G(\mu)\vphi(\bt;z)\cdot \vphi(\bt';-z)}
X(\bt;z)X(\bt;\ld,\mu)\tau(\bt)X(\bt';-z)\tau(\bt') \nn\\
&-\res_z\big(a(z,\ld,\mu)e^{-\xi(\bt;\ld)+\xi(\bt;\mu)}\sqrt{G(-\ld)G(\mu)\vphi(\bt;z)\cdot
\vphi(\bt';-z)}\,\times \nn\\
& \qquad \times
X(\bt;z)G(-\ld)G(\mu)\tau(\bt)X(\bt';-z)\tau(\bt')\big)
\label{Xtau1}
\end{align}
with
\[
a(z,\ld,\mu)=\frac{1-{\mu}/{z}}{1+{\mu}/{z}}\frac{1+{\ld}/{z}}{1-{\ld}/{z}}
-\frac{1-{z}/{\mu}}{1+{z}/{\mu}}\frac{1+{z}/{\ld}}{1-{z}/{\ld}}.
\]
The function $a(z,\ld,\mu)$ can be rewritten as
\begin{equation}\label{}
a(z,\ld,\mu)=\frac{1+{\ld}/{z}}{1+{\mu}/{z}}(z-\mu)\delta(z,\ld)
+\frac{\mu}{\ld}\frac{1-{z}/{\mu}}{1-{z}/{\ld}}(z+\ld)\delta(z,-\mu),
\end{equation}
where
$\delta(z,\ld)=\left(z(1-\ld/z)\right)^{-1}+\left(\ld(1-z/\ld)\right)^{-1}$
is the Dirac delta function such that $\res_z
f(z)\delta(z,\ld)=f(\ld)$ for any Laurent series $f(z)$, see
\cite{Dickey}. Thus by taking $\bt'=\bt$ in \eqref{Xtau1} and using
\eqref{Gvphi} we have
\begin{align}
&\res_z\sqrt{G(-\ld)G(\mu)\vphi(\bt;z)\cdot \vphi(\bt;-z)}
\frac{G(z)X(\ld,\mu)\tau(\bt)}{\tau(\bt)}\frac{G(-z)\tau(\bt)}{\tau(\bt)}
\nn
\\
=&\frac{2(\ld-\mu)}{1+\mu/\ld}e^{-\xi(\bt;\ld)+\xi(\bt;\mu)}\sqrt{G(-\ld)G(\mu)\vphi(\bt;\ld)\cdot
\vphi(\bt;-\ld)}\, \frac{G(\mu)\tau(\bt)}{\tau(\bt)}\frac{G(-\ld)\tau(\bt)}{\tau(\bt)} \nn\\
&
+\frac{\mu}{\ld}\frac{2(\ld-\mu)}{1+\mu/\ld}e^{-\xi(\bt;\ld)+\xi(\bt;\mu)}
\sqrt{G(-\ld)G(\mu)\vphi(\bt;-\mu)\cdot \vphi(\bt;\mu)}\,
\frac{G(-\ld)\tau(\bt)}{\tau(\bt)}\frac{G(\mu)\tau(\bt)}{\tau(\bt)}
\nn\\
=&\frac{2\ld(\ld-\mu)}{\ld+\mu}\sqrt{\frac{G(\mu)\vphi(\bt;-\ld)}{\vphi(\bt;\mu)}}\,w(\bt;\mu)w(\bt;-\ld)
\nn\\
&+\frac{2\mu(\ld-\mu)}{\ld+\mu}\sqrt{\frac{G(-\ld)\vphi(\bt;\mu)}{\vphi(\bt;-\ld)}}\,w(\bt;\mu)w(\bt;-\ld)
\nn\\
=&\frac{2(\ld-\mu)}{\mu+\ld}w(\bt;\mu)w(\bt;-\ld) \left(\ld
G(\mu)\sqrt{\frac{\vphi(\bt;-\ld)}{\vphi(\bt;-\mu)}}+\mu
G(-\ld)\sqrt{\frac{\vphi(\bt;\mu)}{\vphi(\bt;\ld)}}\right).
\label{Xtau2}
\end{align}
Recall $\vphi(\bt;z)=1+O(1/z^2)$, hence the left-hand side of this
equation is
\begin{align}\label{}
\mathrm{l.h.s.}=&\res_z\frac{G(z)X(\ld,\mu)\tau(\bt)}{\tau(\bt)}\frac{G(-z)\tau(\bt)}{\tau(\bt)}
\nn
\\
=&\frac{-2\pd_x\left(X(\ld,\mu)\tau(\bt)\right)}{\tau(\bt)}
+\frac{X(\ld,\mu)\tau(\bt)}{\tau(\bt)}\frac{2\pd_x\tau(\bt)}{\tau(\bt)}
\nn\\
=&-2\pd_x\frac{X(\ld,\mu)\tau(\bt)}{\tau(\bt)}.
\end{align}
Substitute it into \eqref{Xtau2}, then we obtain \eqref{Xtau}. The
lemma is proved.
\end{prf}

One expands the vertex operator \eqref{X} formally as
\begin{equation}\label{}
{X}(\ld,
\mu)=\sum_{m=0}^\infty\frac{(\mu-\ld)^m}{m!}\sum_{l=-\infty}^\infty
\ld^{-m-l}{W}_l^{(m)}
\end{equation}
with
\[
{W}_l^{(m)}=\res_\ld\left(\ld^{m+l-1}\pd_\mu^m|_{\mu=\ld}{X}(\ld,
\mu)\right).
\]
For instance, it is straightforward to calculate
\begin{align*}
&W_l^{(0)}=\dt_{l0}, \quad W_l^{(1)}=p_l, \quad
W_l^{(2)}=\sum_{i+j=l}:p_i p_j:-(l+1)p_l, \\
&W_l^{(3)}=\sum_{i+j+k=l}:p_i p_j
p_k:-\frac{3}{2}(l+2)\sum_{i+j=l}:p_i p_j:+(l+2)(l+1) p_l.
\end{align*}
Here for convenience we assume $p_i=0$ for even $i$.

\begin{prp}\label{thm-taus}
For the CKP hierarchy, the additional symmetries \eqref{sml} acting
on the tau function are given by the following formula
\begin{align}\label{taus}
&\sum_{m=0}^\infty\frac{(\mu-\ld)^m}{m!}\sum_{l=-\infty}^\infty
\ld^{-m-l-1}\pd_x\left(\frac{1}{\tau(\bt)}\frac{\pd\tau(\bt)}{\pd
s_{m,m+l}}-\frac{1}{m+1}\frac{W_l^{(m+1)}\tau(\bt)}{\tau(\bt)}\right)
\nn \\
=&w(\bt;\mu)w(\bt;-\ld) \left(1-\frac{\ld}{\mu+\ld}
G(\mu)\sqrt{\frac{\vphi(\bt;-\ld)}{\vphi(\bt;-\mu)}}-\frac{\mu}{\mu+\ld}
G(-\ld)\sqrt{\frac{\vphi(\bt;\mu)}{\vphi(\bt;\ld)}}\right).
\end{align}
\end{prp}
\begin{prf}
Let
\begin{equation}\label{}
Z(\ld,\mu)=\frac{1}{\mu-\ld}(X(\ld,\mu)-1)
=\sum_{m=0}^\infty\frac{(\mu-\ld)^m}{m!}\sum_{l=-\infty}^\infty
\ld^{-m-l-1}\frac{W_l^{(m+1)}}{m+1}.
\end{equation}
Lemma~\ref{thm-Xtau} implies
\begin{align}
&\pd_x\frac{Z(\ld,\mu)\tau(\bt)}{\tau(\bt)} \nn\\
=&w(\bt;\mu)w(\bt;-\ld) \left(\frac{\ld}{\mu+\ld}
G(\mu)\sqrt{\frac{\vphi(\bt;-\ld)}{\vphi(\bt;-\mu)}}+\frac{\mu}{\mu+\ld}
G(-\ld)\sqrt{\frac{\vphi(\bt;\mu)}{\vphi(\bt;\ld)}}\right).
\end{align}
Hence to show \eqref{taus} we only need to verify
\begin{align}\label{logtaus}
&\sum_{m=0}^\infty\frac{(\mu-\ld)^m}{m!}\sum_{l=-\infty}^\infty
\ld^{-m-l-1}\pd_x\frac{\pd\log\tau(\bt)}{\pd
s_{m,m+l}}=w(\bt;\mu)w(\bt;-\ld).
\end{align}
In fact, according to \eqref{sml} and \eqref{a1tau} one has
\[
\res A_{m,m+l}=-\frac{\pd a_1}{\pd
s_{m,m+l}}=2\pd_x\frac{\pd\log\tau(\bt)}{\pd s_{m,m+l}}.
\]
Thus the equality \eqref{logtaus} is deduced by taking the residue
of the operator \eqref{Yw}. Therefore the proposition is proved.
\end{prf}

\begin{cor}\label{thm-taus01}
The additional symmetries \eqref{sml} with $m=0$ and $1$ can be
represented as follows
\begin{itemize}
\item[(i)] for $l\in\Z$,
\begin{equation}\label{}
\frac{\pd\tau}{\pd
s_{0l}}=\left(W_l^{(1)}+\dt_{l0}\,c^{(1)}\right)\tau;
\end{equation}
\item[(ii)] for $i\in\Z$,
\begin{align}\label{taus2i}
&\frac{\pd\tau}{\pd s_{1,2i}}=\frac{1}{2}W_{2i-1}^{(2)}\tau;
\\
&\frac{\pd\tau}{\pd
s_{1,2i+1}}-\left(\frac{1}{2}W_{2i}^{(2)}+\dt_{i0}\,c^{(2)}\right)\tau
=\left\{\begin{array}{cl}
          0, & i\le0; \\
          \tau\pd_x^{-1}T_{2i+1}, & i\ge1,
        \end{array}
 \right. \label{taus2i1}
\end{align}
in which
\begin{equation}\label{T}
T_{2i+1}=-\frac{1}{2}\res_\ld
\ld^{2i+1}\frac{G(\ld)\tau}{\tau}\frac{G(-\ld)\tau}{\tau}
\sqrt{\frac{\vphi(\bt;-\ld)}{\vphi(\bt;\ld)}}(N(\ld)-\pd_\ld)\vphi(\bt;\ld)
\end{equation}
with
\[
N(\ld)=\sum_{k\in\Zop}\frac{2}{\ld^{k+1}}\frac{\pd}{\pd t_k}.
\]
\end{itemize}
Here $c^{(1)}$ and $c^{(2)}$ are certain constants that arise from a
central extension of the $w^C_{\infty}$-algebra.
\end{cor}
\begin{prf}
Clearly the right-hand side of \eqref{taus} vanishes whenever
$\mu=\ld$, which implies the first assertion. Let us proceed to show
the second one.

It is easy to see
\[
[\pd_\ld, G(\ld)]=N(\ld)G(\ld), \quad [\pd_\ld,
G(-\ld)]=-N(\ld)G(-\ld).
\]
Denote
\begin{equation}\label{}
\chi(\bt;\ld,\mu)=1-\frac{\ld}{\mu+\ld}
G(\mu)\sqrt{\frac{\vphi(\bt;-\ld)}{\vphi(\bt;-\mu)}}-\frac{\mu}{\mu+\ld}
G(-\ld)\sqrt{\frac{\vphi(\bt;\mu)}{\vphi(\bt;\ld)}}.
\end{equation}
One has $\chi(\bt;\ld,\ld)=0$ and
\begin{align}\label{}
&\left.\pd_\mu\right|_{\mu=\ld}\chi(\bt;\ld,\mu)
\nn\\
=&\frac{1}{4}G(\ld)\pd_\ld\log
\vphi(\bt;-\ld)-\frac{1}{4}G(-\ld)\pd_\ld\log \vphi(\bt;\ld) \nn
\\
=&\frac{1}{4}(-N(\ld)+\pd_\ld)\log
\vphi(\bt;\ld)-\frac{1}{4}(N(\ld)+\pd_\ld)\log \vphi(\bt;-\ld) \nn
\\
=&-\frac{1}{4}N(\ld)\log \left(\vphi(\bt;\ld)\vphi(\bt;-\ld)\right)
+ \frac{1}{4}\pd_\ld\log\frac{\vphi(\bt;\ld)}{\vphi(\bt;-\ld)}.
\label{chi1}
\end{align}
Hence
\begin{align}
&\pd_x\left(\frac{1}{\tau(\bt)}\frac{\pd\tau(\bt)}{\pd
s_{1,l+1}}-\frac{1}{2}\frac{W_l^{(2)}\tau(\bt)}{\tau(\bt)}\right)
\nn\\
=&\res_\ld\ld^{l+1}\left.\pd_\mu\right|_{\mu=\ld}\left(w(\bt;\mu)w(\bt;-\ld)\chi(\bt;\ld,\mu)\right)
\nn\\
=&-\frac{1}{4}\res_\ld\ld^{l+1}w(\bt;\ld)w(\bt;-\ld)\left(N(\ld)\log\left(\vphi(\bt;\ld)\vphi(\bt;-\ld)\right)
-\pd_\ld\log\frac{\vphi(\bt;\ld)}{\vphi(\bt;-\ld)}\right).\label{taus1}
\end{align}
Since \eqref{chi1} is an even function in $\ld$, then \eqref{taus1}
vanishes whenever $l$ is odd. Namely, the equality \eqref{taus2i} is
verified. Recall $\vphi(\bt;\ld)=1+O(1/\ld^2)$, then \eqref{taus1}
vanishes when $l\le0$. If $l=2i$ is even, we simplify the right hand
side of \eqref{taus1} to
\[
-\frac{1}{2}\res_\ld\ld^{2 i+1}w(\bt;\ld)w(\bt;-\ld)\left(N(\ld)\log
\vphi(\bt;\ld) -\pd_\ld\log \vphi(\bt;\ld)\right),
\]
and substitute into it with
\[
w(\bt;\ld)=\sqrt{\vphi(\bt;\ld)}\,\frac{G(\ld)\tau(\bt)}{\tau(\bt)}e^{\xi(\bt;z)}
\]
(see \eqref{wavef} and \eqref{phitau}), thus we recast it to
$T_{2i+1}$ and verify \eqref{taus2i1}. The corollary is proved.
\end{prf}

\begin{exa}\label{exa=T}
Let us illustrate how to compute $T_{2i+1}$ in the tails of
\eqref{taus2i1}. For this purpose, we expand
\[
G(\ld)=1+\sum_{j\ge1}\sg_j(-\tilde{\bm{\pd}})\frac{1}{\ld^j}, \quad
\tilde{\bm{\pd}}=\left(\frac{2}{1}\frac{\pd}{\pd t_1}, 0,
\frac{2}{3}\frac{\pd}{\pd t_3}, 0, \dots\right),
\]
where $\sg_j$ are elementary Schur polynomials defined by
\[
\exp\left(\sum_{i\ge1}q_i
z^i\right)=\sum_{j\ge0}\sg_j(q_1,q_2,\dots)z^j.
\]
Denote $\ta=\log\tau$, and $\ta_x=\pd_x\ta$,
$\ta_{x\,t_3}=\pd_x\pd_{t_3}\ta$ etc. (recall $x=t_1$). One has
\begin{align}\label{}
&\vphi(\bt;\ld)=1+\frac{1}{\ld}(G(\ld)-1)\ta_x
=1+\sum_{j\ge1}\frac{1}{\ld^{j+1}}\sg_j(-\tilde{\bm{\pd}})\ta_x=1+O(1/\ld^2),
\\
&(N(\ld)-\pd_\ld)\vphi(\bt;\ld)\nn\\
=&\frac{1}{\ld^2}(G(\ld)-1)\ta_x-\frac{1}{\ld}N(\ld)\ta_x
\nn\\
=&\sum_{i\ge1}\left(\frac{1}{\ld^{2i+1}}\left(\sg_{2i-1}(-\tilde{\bm{\pd}})-2\frac{\pd}{\pd
t_{2i-1}}\right)\ta_x+\frac{1}{\ld^{2i+2}}\sg_{2i}(-\tilde{\bm{\pd}})\ta_x\right)=O(1/\ld^3),
\\
&\frac{G(\ld)\tau}{\tau}\frac{G(-\ld)\tau}{\tau}
\nn\\
=&\exp\left((G(\ld)+G(-\ld)-2)\ta\right)
\nn\\
=&\exp\left(2\sum_{i\ge1}\frac{1}{\ld^{2i}}\sg_{2i}(-\tilde{\bm{\pd}})\ta\right)=1+O(1/\ld^2).
\end{align}
Substitute them into \eqref{T}, then it is straight forward to
obtain
\begin{align}\label{T35}
T_3=-\ta_{xxx}, \quad 
T_5=-\pd_x\left(\frac{2}{3}\ta_{x\,t_3}+\frac{1}{3}\ta_{x x x
x}+4\ta_{x x}^2\right).
\end{align}
Generally, for $i\ge1$,
\begin{equation}\label{Tail}
T_{2i+1}=-\frac{1}{2}\sg_{2i}(-\tilde{\bm{\pd}})\ta_x+\hbox{
nonlinear terms in derivatives of $\ta$}.
\end{equation}
Note that the nonlinear terms are trivial whenever $i=1$.

\begin{lem}\label{thm-T}
For every $i\ge1$, it holds that $T_{2i+1}=\pd_x\tilde{T}_{2i}$ for some polynomial 
\begin{equation}\label{}
\tilde{T}_{2i}\in\C\left[\frac{\pd^{m_1+\dots+m_s}}{\pd
t_{k_1}^{m_1}\dots\pd t_{k_s}^{m_s}}\ta;~ m_1+\dots+m_s\ge2\right].
\end{equation}
Moreover, each polynomial $\tilde{T}_{2i}$ is homogeneous of degree
$2 i$ if we assign
\[
\deg \frac{\pd^{m_1+\dots+m_s}}{\pd t_{k_1}^{m_1}\dots\pd
t_{k_s}^{m_s}}\ta =m_1 k_1+\dots+m_s k_s.
\]
\end{lem}
The validity of this lemma will be verified below, though not in so
direct a way.
\end{exa}

With the same method as in Corollary~\ref{thm-taus01}, from
\eqref{taus} one can calculate $\pd\tau/\pd s_{m,m+l}$ for $m\ge2$.
However, the formulae turn out complicated.

Now we conclude that, for the CKP hierarchy, when lifting the
actions of additional symmetries on the wave function to the actions
on the tau function, one has not only a central extension of the
$w_\infty^C$-algebra but also certain nontrivial ``tails'' like
given by $T_{2i+1}$. This is the main difference between the
additional symmetries for the CKP hierarchy and those as for the KP
hierarchy considered before \cite{ASvM, Di, vdL, Tu, Wu}. For what
reason the CKP hierarchy is so special is not clear yet.

\section{Virasoro symmetries for Drinfeld-Sokolov hierarchies of type C}

In the celebrated work \cite{DS}, Drinfeld and Sokolov associated an
integrable hierarchy of KdV type to each affine Kac-Moody algebra
$\fg$. Instead of getting into their general construction, in this
section we only consider the case that $\fg$ is of type $C_n^{(1)}$.
More exactly, we want to apply the above results for the CKP
hierarchy to study the Virasoro symmetries for the Drinfeld-Sokolov
hierarchy of type $C_n^{(1)}$. Our inspiration is from the reduction
relation of the Lie algebras $C_\infty\to C_n^{(1)}$ underlying
these hierarchies \cite{JM83, Kac}.

\subsection{Drinfeld-Sokolov hierarchies of type C}

For any integer $n\ge2$ fixed, the simple Lie algebra
$\mathring{\fg}$ of type $C_n$ can be realized as
\[
\mathring{\fg}=\mathfrak{sp}(2n)=\set{A\in\C^{2n\times 2n}\mid A=-S
A^T S^{-1} }
\]
where $S=\diag(1,-1,1,-1,\dots,1,-1)$ and the superscript ``$T$''
means the transpose relative to the secondary diagonal \cite{DS}.
Then one realizes the Kac-Moody algebra $\fg$ of affine type
$C_n^{(1)}$ as
\[
\fg=\mathring{\fg}[\ld,1/\ld]\oplus\C\,c\oplus\C\,d
\]
 with $c$ being
the canonical central element and $d$ the scaling element. In more
details, let $e_{i,j}$ be the $2n\times 2n$ matrix whose
$(i,j)$-entry takes value $1$ and any other entry vanishes, then a
set of Weyl generators of $\fg$ can be chosen as follows \cite{DS,
Kac}:
\begin{align}\label{efhh}
&e_i=e_{i+1,i}+e_{2n-i+1,2n-i}\quad (1\le i\le n-1),  \\
&e_n=e_{n+1,n},\quad e_0=\ld e_{1,2n}, \\
&f_i=e_{i,i+1}+e_{2n-i,2n-i+1}\quad (1\le i\le n-1),  \\
&f_n=e_{n,n+1},\quad f_0=\ld^{-1} e_{2n,1},\\
&\al^\vee_i=[e_i,f_i] \quad (1\le i\le n), \\
&\al^\vee_0=e_{1,1}-e_{2n,2n}+c.
 \label{efht}
\end{align}

Denote $\Ld=\sum_{i=0}^n e_i$. The elements $\Ld^{k}$ with
$k\in\Z^{\mathrm{odd}}$ generate the principal Heisenberg subalgebra
of $\fg$. Moreover, they satisfy
\[
(\Ld^{k}\mid\Ld^{l})=2\,n\,\dt_{k,-l}, \quad k,l\in\Z^{\mathrm{odd}}
\]
for the standard invariant bilinear (Killing) form
$(\,\cdot\mid\cdot)$ on $\fg$. Note that $2\,n$ is the Coxeter
number.

Introduce a matrix operator
\begin{equation}\label{msL}
\sL= D+\Ld + q
\end{equation}
with $D={\od}/{\od x}$ and $q$ being a smooth function of $x$ that
takes value in the Borel subalgebra of $\mathring{\fg}$ generated by
$\al_i^\vee$ and $f_i$ with $1\le i\le n$. The nilpotent subalgebra,
say $\mathfrak{n}$, generated by $f_i$ with $1\le i\le n$, induces a
group of gauge transformations of $\sL$ as
\begin{equation}\label{gauge}
\sL\mapsto e^{\ad_N}\sL,  \quad N\in \mathfrak{n}.
\end{equation}

The Drinfeld-Sokolov hierarchy associated to $\fg$ is defined to be
\begin{equation}\label{dnm}
\frac{\pd \sL}{\pd t_k}=[\mathscr{A}(\Ld^k), \sL],  \quad k\in \Zop
\end{equation}
modulo the gauge transformations \eqref{gauge}. Here
$\mathscr{A}(\Ld^k)$, depending on $\Ld^k$, are certain $\fg$-valued
differential polynomials in $q$ such that the right hand side of
\eqref{dnm} takes value in the Borel subalgebra of $\mathring{\fg}$,
see \cite{DS} (also \cite{Wu-DS}) for details.

For the equivalence class of $\sL$ with respect to the
transformations \eqref{gauge}, a representative element can be
chosen as
\begin{align}\label{qcan}
\sL^{\mathrm{can}}=D+\Ld+q^{\mathrm{can}}, \quad
q^{\mathrm{can}}=-\sum_{i=1}^{n}\frac{u_i}{2}(e_{1,2i}+e_{2n-2i+1,
2n})
\end{align}
with scalar functions $u_i$. According to the theory of \cite{DS},
the canonical operator \eqref{qcan} induces a scalar differential
operator
\begin{equation}\label{cL}
\cL=D^{2n}+\frac{1}{2}\sum_{i=1}^n\left( (u_i+r_i )
D^{2n-2i}+D^{2n-2i}(u_i+r_i )\right),
\end{equation}
where $r_i=r_i(u_1,\dots,u_{i-1})$ are differential polynomials in
their arguments and particularly $r_1=0$. Hence the hierarchy
\eqref{dnm} is converted to the following system of Lax equations
\begin{equation}\label{cLt}
\frac{\pd\cL}{\pd t_k}=\left[ (\cL^{k/2n})_+, \cL\right], \quad
k\in\Zop.
\end{equation}

The Drinfeld-Sokolov hierarchy  \eqref{dnm} of type $C_n^{(1)}$
carries a bi-Hamiltonian structure \cite{DS}. In \cite{Wu-DS} a set
of Hamiltonian densities were selected appropriately such that they
define a tau function, say, $\tilde{\tau}$ (to be distinguished from
the previous notation $\tau$ of the CKP hierarchy). With the same
method as in \cite{Wu-DS} (see equation~(5.13) there), we have
\begin{equation}\label{tautd}
\pd_x^2\log\tilde\tau=\frac{(-\Ld\mid
q^{\mathrm{can}})}{(\Ld\mid\Ld^{-1})}=\frac{u_1}{2\,n}.
\end{equation}

\begin{thm}[\cite{Wu-DS}]
The Virasoro symmetries for the Drinfeld-Sokolov hierarchy
\eqref{dnm}, i.e., \eqref{cLt}, can be represented as
\begin{equation}\label{tauvir}
\frac{\pd\tilde\tau}{\pd\beta_j}=V_j\tilde\tau+\tilde\tau\,O_j,
\quad j=-1,0,1,2,\dots.
\end{equation}
Here
\begin{equation}\label{V}
V_j=\frac{1}{4n}\sum_{k\in\Z^{\mathrm{odd}}}:\tilde{p}_k\tilde{p}_{2n\,j-k}:
+\dt_{j0}\,c_n
\end{equation}
with
\[
\tilde{p}_k=\frac{\pd}{\pd t_k}, \quad \tilde{p}_{-k}=k\,t_k, \qquad
k\in\Zop,
\]
and  $c_n$ being a constant; the terms $O_j$ are differential
polynomials in second-order derivatives of $\log\tilde\tau$ with
respect to the time variables, and, in particular, $O_{-1}=O_0=0$.
\end{thm}

We write the operators $V_j$ explicitly as
\begin{align}\label{}
V_{-1}=&\frac{1}{2\,n}\sum_{k\in\Zop}(k+2\,n)t_{k+2n}\frac{\pd}{\pd
t_k}+\frac{1}{4\,n}\sum_{k+l=2n}k\,l\,t_k\,t_l, \\
V_{0}=&\frac{1}{2\,n}\sum_{k\in\Zop} k\,t_{k}\frac{\pd}{\pd
t_k}+c_n, \\
V_{j}=&\frac{1}{4\,n}\sum_{k+l=2n j}\frac{\pd^2}{\pd t_k\pd
t_l}+\frac{1}{2\,n}\sum_{k\in\Zop}k\,t_{k}\frac{\pd}{\pd t_{k+2 n
j}}, \quad j\ge1,
\end{align}
where all indices $k$ and $l$ lie in $\Zop$. Choose
\begin{equation}\label{cn}
c_n=\frac{n}{24}\left(1+\frac{1}{2\,n^2}\right),
\end{equation}
then $V_j$ satisfy the Virasoro commutation relation (see, for
example, \cite{KR})
\begin{equation}\label{}
[V_i, V_j]=(i-j)V_{i+j}, \quad i, j\ge-1.
\end{equation}

The terms $O_j$ in \eqref{tauvir} are called obstacles in
linearizing Virasoro symmetries in \cite{Wu-DS}. The Virasoro
symmetries are said to be linearized if all such $O_j$ vanish, which
is a crucial property of an integrable hierarchy of topological type
\cite{DZ}. We remark that all Drinfeld-Sokolov hierarchies
associated to ADE-type affine Kac-Moody algebras, either untwisted
or twisted, possess linearized Virasoro symmetries \cite{Wu-DS}, see
also \cite{DZ, HMSG, Wu}. However, for the Drinfeld-Sokolov
hierarchies of type C, it was unknown whether these obstacles $O_j$
vanish or not, since it is not easy to compute them starting from
the original definition in \cite{Wu-DS}. Such obstacles will be
calculated alternatively in the forthcoming subsection in
consideration of that \eqref{cLt} is indeed a subhierarchy of the
CKP hierarchy \eqref{ckp}.

\subsection{Non-linearizable Virasoro symmetries}

Given an integer $n\ge2$, unless otherwise stated the
pseudo-differential operator \eqref{Lckp} is henceforth assumed to
satisfy
\begin{equation}\label{L2nm}
(L^{2n})_-=0.
\end{equation}
Under this constraint, the CKP hierarchy \eqref{ckp} is reduced to
the hierarchy \eqref{cLt} with $\cL=L^{2n}$, and the bilinear
equation \eqref{blckp} becomes
\begin{equation}\label{blcn}
\res_{z} z^{2n j} w(\mathbf{t};z)w(\mathbf{t}';-z)=0, \quad j\ge0.
\end{equation}
Meanwhile, the Poisson brackets \eqref{pb1} and \eqref{pb2} admit
the constraint \eqref{L2nm}, hence one rederives the bi-Hamiltonian
structure for the Drinfeld-Sokolov hierarchy achieved in \cite{DS}.
The Hamiltonians are also given by the formulae \eqref{Hk}, thus the
tau function $\tau$ of the CKP hierarchy can be reduced to a tau
function of the hierarchy \eqref{cLt}.

\begin{prp}
For the Drinfeld-Sokolov hierarchy \eqref{cLt} of type $C_n^{(1)}$,
the tau functions $\tau$ reduced from that of the CKP hierarchy and
$\tilde\tau$ as recalled in the preceding subsection satisfy
$\tau^2=\tilde\tau$ (up to a factor of the form $\exp\left(\sum c_k
t_k\right)$ with constant $c_k$).
\end{prp}
\begin{prf}
The proof is similar to that of Propositions~5.2 and 5.4 in
\cite{Wu-DS} for Drinfeld-Sokolov hierarchies of types A and D.
According to \eqref{tau}, \eqref{cL} and \eqref{tautd}, we have
\[
\pd_x^2\log\tau^2=\res\,L=\res\,\cL^{1/2n}=\frac{u_1}{2\,n}=\pd_x^2\log\tilde\tau.
\]
Hence
\[
\pd_x^2\left(\frac{\pd^2\log\tau^2}{\pd t_k\pd
t_l}-\frac{\pd^2\log\tilde\tau}{\pd t_k\pd t_l}\right)=0, \quad k,
l\in\Zop.
\]
Note that the terms in parentheses are differential polynomials in
the coefficients of $\cL$, namely, in $(u_1, u_2, \dots, u_n)$,
hence their difference vanishes indeed. It follows that $\tau^2$ and
$\tilde\tau$ coincide. The proposition is proved.
\end{prf}

Part of the additional symmetries \eqref{sml} for the CKP hierarchy
are compatible with the constraint \eqref{L2nm}. In fact, for
$j\ge-1$, one has
\begin{align}
\left(\frac{\pd L^{2n}}{\pd s_{1,2n j+1}}\right)_-=&[-(A_{1,2 n
j+1})_-,L^{2n}]_- \nn\\
=&[-A_{1,2 n j+1},L^{2n}]_- \nn\\
=&4n(L^{2n(j+1)})_-=0.
\end{align}
Denote $s_j=4\,n\,s_{1,2 n j+1}$, then
\begin{equation}\label{}
\frac{\pd\cL}{\pd s_j}=\frac{1}{4\,n}[-(A_{1,2 n j+1})_-,\cL], \quad
j\ge-1
\end{equation}
are symmetries for the reduced hierarchy \eqref{cLt}. Moreover,
Proposition~\ref{thm-st} implies that these symmetries satisfy the
Virasoro commutation relation
\begin{equation}\label{}
\left[\frac{\pd}{\pd s_i},\frac{\pd}{\pd
s_j}\right]=(j-i)\frac{\pd}{\pd s_{i+j}}, \quad i,j\ge-1
\end{equation}
when acting on $\cL$, or, on the dressing operator $\Phi$ given as
in \eqref{Phi}. According to \eqref{taus2i1} we have
\begin{equation}\label{tauvir2}
\frac{\pd\tau}{\pd s_{j}}=L_j\tau+\frac{\tau}{4\,n}\pd_x^{-1}T_{2n
j+1}, \quad j\ge-1,
\end{equation}
where
\begin{equation}\label{}
L_j=\frac{1}{8\,n}W_{2n j}^{(2)}+\dt_{j 0}\,\frac{c_n}{2}.
\end{equation}
Explicitly, one has
\begin{align}\label{}
L_{-1}=&\frac{1}{2\,n}\sum_{k\in\Zop}(k+2\,n)t_{k+2n}\frac{\pd}{\pd
t_k}+\frac{1}{8\,n}\sum_{k+l=2n}k\,l\,t_k\,t_l, \\
L_{0}=&\frac{1}{2\,n}\sum_{k\in\Zop} k\,t_{k}\frac{\pd}{\pd
t_k}+\frac{c_n}{2}, \\
L_{j}=&\frac{1}{2\,n}\sum_{k+l=2n j}\frac{\pd^2}{\pd t_k\pd
t_l}+\frac{1}{2\,n}\sum_{k\in\Zop}k\,t_{k}\frac{\pd}{\pd t_{k+2 n
j}}, \quad j\ge1.
\end{align}
Here the constant $c_n/2$ (see \eqref{cn}) in $L_0$ is chosen for
the validity of the following theorem.

\begin{thm}
For the Drinfeld-Sokolov hierarchy \eqref{cLt} of type $C_n^{(1)}$,
the Virasoro symmetries \eqref{tauvir2} and \eqref{tauvir} coincide.
More precisely, acting on the tau function $\tilde\tau$ it holds
that
\begin{align}\label{}
\frac{\pd\tilde\tau}{\pd s_j}=\frac{\pd\tilde\tau}{\pd \beta_j},
\quad j\ge-1.
\end{align}
\end{thm}
\begin{prf}
For $j\ge-1$, we write $L_j=L_j^{(2)}+L_j^{(1)}+L_j^{(0)}$, where
$L_j^{(\nu)}$ is the part of the $\nu$th order derivations in $L_j$.
For instance,
\[
L_0^{(0)}=\frac{c_n}{2}, \quad
L_1^{(2)}=\frac{1}{2\,n}\sum_{k+l=2n}\frac{\pd^2}{\pd t_k\pd t_l}.
\]
Similarly we write $V_j=V_j^{(2)}+V_j^{(1)}+V_j^{(0)}$ for $V_j$
given in \eqref{V}. It is easy to see
\[
L_j^{(2)}=2\,V_j^{(2)}, \quad L_j^{(1)}=V_j^{(1)}, \quad
L_j^{(0)}=\frac{1}{2}V_j^{(0)}.
\]

Since $\tilde\tau=\tau^2$, then
\begin{align*}
V_j\tilde\tau=&4\tau V_j^{(2)}\tau+2\tau V_j^{(1)}\tau+ \tau
V_j^{(0)}\tau -2\tau^2 V_j^{(2)}\log\tau \nn\\
=&2\tau L_j\tau -\tilde\tau V_j^{(2)}\log\tilde\tau.
\end{align*}
Comparing \eqref{tauvir2} and \eqref{tauvir}, we have
\begin{align}\label{}
&\pd_x\left(\frac{\pd\log\tilde\tau}{\pd
s_j}-\frac{\pd\log\tilde\tau}{\pd \beta_j}\right) \nn\\
=&\pd_x\left(\frac{2}{\tau}\left(L_j\tau+\frac{\tau}{4\,n}\pd_x^{-1}T_{2n
j+1}\right)-\frac{1}{\tilde\tau}(V_j\tilde\tau+\tilde\tau O_j)\right) \nn\\
=&\frac{1}{2\,n} T_{2n
j+1}-\pd_x\left(O_j-V_j^{(2)}\log\tilde\tau\right). \label{sbt1}
\end{align}
The left-hand side depends linearly on $\log\tilde\tau$, so does the
right-hand side. Observe \eqref{Tail} and recall that $O_j$ are
differential polynomials in second-order derivatives of
$\log\tilde\tau$, then the right-hand side of \eqref{sbt1} must be
of the form $\pd_x R_j\log\tilde\tau$ for some linear operator
$R_j\in\C[\pd/\pd t_1, \pd/\pd t_3, \dots]$. Thus acting on
$\log\tilde\tau$ one has
\[
\frac{\pd}{\pd s_j}=\frac{\pd}{\pd \beta_j}+R_j, \quad j\ge0,
\]
where $R_{-1}=R_{0}=0$. In fact, all $R_j$ must vanish by virtue of
the Virasoro commutation relations for the symmetries $\pd/\pd s_j$
and for $\pd/\pd\beta_j$ respectively. Therefore
\begin{equation}\label{}
\frac{\pd\log\tilde\tau}{\pd s_j}=\frac{\pd\log\tilde\tau}{\pd
\beta_j}, \quad j\ge-1.
\end{equation}
The theorem is proved.
\end{prf}

From the proof we also know that each $T_{2n j+1}$ is a total
derivatives of some differential polynomial in second-order
derivatives of $\log\tilde\tau$ with respect to the time variables.
Hence we obtain an alternative representation for the obstacles that
were introduced from Kac-Moody-Virasoro algebra in \cite{Wu-DS}.
\begin{cor}\label{thm-O}
The obstacles $O_j$ in \eqref{tauvir} can be represented as
\begin{equation}\label{OT}
O_j=\frac{1}{2\,n}\left(\pd_x^{-1} T_{2n
j+1}+\frac{1}{2}\sum_{k+l=2n j}\frac{\pd^2\log\tilde\tau}{\pd t_k\pd
t_l}\right), \quad j\ge1,
\end{equation}
where $T_{2n j+1}$ are given in \eqref{T} with
$\tau=\exp\left(\frac1{2}\log\tilde{\tau}\right)$.
\end{cor}

\begin{cor}\label{thm-Onon}
For the Drinfeld-Sokolov hierarchy of type $C_n^{(1)}$ with $n\ge2$,
the obstacles $O_j\ne0$ when $j\ge1$.
\end{cor}
\begin{prf}
Substitute \eqref{Tail} into \eqref{OT}, then the part linear in
$\log\tilde\tau$ of $O_j$ with $j\ge1$ is
\begin{align}
O_j^{(1)}=&\frac{1}{4\,n}\left(-\frac{1}{2}\sg_{2 n
j}(-\tilde{\bm{\pd}})+\sum_{k+l=2n j}\frac{\pd^2}{\pd t_k\pd
t_l}\right)\log\tilde\tau \nn\\
=&\frac{1}{4\,n}\sum_{k+l=2n
j}\left(1-\frac{1}{k\,l}\right)\frac{\pd^2\log\tilde\tau}{\pd t_k\pd
t_l} \nn\\
&-\frac{1}{8\,n}\sum_{\scriptsize\hbox{$\begin{array}{c} k_1
m_1+\dots+k_r m_r=2 n j \\m_1+\dots+m_r\ge3; k_1<\dots<k_r
\end{array}$}}\left(\prod_{\nu=1}^{r}\frac{1}{m_\nu!}\left(\frac{2}{k_\nu}
\frac{\pd}{\pd t_{k_\nu}}\right)^{m_\nu}\right)\log\tilde\tau.
\end{align}
In particular, taking $j=1$ one derives $O_1^{(1)}\ne0$ hence
$O_1\ne0$. Here it is adopted the fact that the flows $\pd/\pd t_k$
with $k=1, 3, \dots, 2\,n-1$ in the hierarchy \eqref{cLt} are
independent so that the linear part $O_1^{(1)}$ cannot be canceled
by the omitted nonlinear part (cf. Remark~\ref{rmk-kdv} below).

Furthermore, provided $O_j=0$ for some $j>1$, it follows that
$O_{j-1}=0$ due to the commutation relation between $\pd/\pd s_{-1}$
and $\pd/\pd s_j$. Then step by step one deduces $O_1=0$, which is a
contradiction. Therefore the theorem is proved.
\end{prf}

\vskip 2ex

\begin{prfof}{Theorem~\ref{thm-dsc}}
The theorem is a combination of Corollaries~\ref{thm-O} and
\ref{thm-Onon}.
\end{prfof}

\begin{rmk}\label{rmk-kdv}
The condition $n\ge2$ in Corollary~\ref{thm-Onon} is essential.
Otherwise, suppose $n=1$, then the reduced hierarchy \eqref{cLt}
with
\[
\cL=D^2+u
\]
is nothing but the KdV hierarchy, or equivalently, the
Drinfeld-Sokolov hierarchy associated to the affine Kac-Moody
algebra of type $A_1^{(1)}$. As it is known, the Virasoro symmetries
for the KdV hierarchy is linearizable (see, for example, \cite{dV,
DS, Wu-DS}).

In fact, according to \eqref{OT} and \eqref{T35}, one has
\begin{align}\label{}
O_1=&\frac{1}{2}\pd_x^{-1}T_3+\frac{1}{4}\pd_x^2\log\tilde\tau=-\frac{1}{4}\pd_x^2\log\tilde\tau
+\frac{1}{4}\pd_x^2\log\tilde\tau=0, \\
O_2=&\frac{1}{2}\pd_x^{-1}T_5+\frac{1}{2}\frac{\pd^2\log\tilde\tau}{\pd x\pd t_3} \nn\\
=&\frac{1}{3}\frac{\pd^2\log\tilde\tau}{\pd x\pd t_3}-
\frac{1}{12}\frac{\pd^4\log\tilde\tau}{\pd x^4}-
\frac{1}{2}\left(\frac{\pd^2\log\tilde\tau}{\pd x^2} \right)^2.
\end{align}
But the function $u=2\pd_x^2\log\tilde\tau$ satisfies the KdV
equation
\[
\frac{\pd u}{\pd t_3}=\frac{1}{4}\frac{\pd^3 u}{\pd x^3}+
\frac{3}{2}u\frac{\pd u}{\pd x}.
\]
One rewrites this equation in term of $\log\tilde\tau$ then achieves
$O_2=0$. Furthermore, the Virasoro commutation relation for the
symmetries $\pd/\pd s_j$ implies $O_j=0$ for all $j\ge3$. Thus the
linearization of Virasoro symmetries for the KdV hierarchy is
derived again, which agrees with the result in the literature.
\end{rmk}

\begin{prfof}{Lemma~\ref{thm-T}}
The independence of the above flows $\pd/\pd t_k$ with $k=1, 3,
\dots, 2\,n-1$ implies that, the term $T_{2n+1}$ for the reduced
hierarchy \eqref{cLt} has the same expression with that for the CKP
hierarchy \eqref{ckp}. Thus $T_{2n+1}$ in the latter case is also a
total derivative of differential polynomial in second-order
derivatives of $\log\tilde\tau=2\log\tau$ with respect to $t_1, t_3,
\dots, t_{2n-1}$. Since $n$ can be arbitrarily chosen, then such a
property is possessed by every $T_{2 i+1}$ with $i\ge1$ for the CKP
hierarchy. In other words,  $T_{2 i+1}$ is a total derivative of
polynomial in at-least-second-order derivatives of $\log\tau$ with
respect to the time variables. The homogeneity of $T_{2 i+1}$ is
easily observed from the definition \eqref{T}. Therefore
Lemma~\ref{thm-T} is proved.
\end{prfof}

\section{Conclusion}

We have defined a single tau function of the CKP hierarchy from its
Hamiltonian densities. With the help of this tau function, the CKP
hierarchy is represented into a generalized Hirota bilinear equation
\eqref{taubl}, the form of which is different from that of the KP or
of the BKP hierarchy. Furthermore, we have shown that the actions on
the tau function by additional symmetries involve strictly more than
a central extension of the $w^C_\infty$-algebra. It is interesting
to develop similar skills to deal with the generalization
\cite{vdLOS} of the CKP hierarchy that contains both normal and
super variables. An answer to it must enrich our knowledge of
integrable hierarchies and their applications.

By reducing additional symmetries for the CKP hierarchy, the
Virasoro symmetries for the Drinfeld-Sokolov hierarchy associated to
affine Kac-Moody algebra of type $C_n^{(1)}$ with $n\ge2$ are
rederived. The Virasoro symmetries coincide with those constructed
in \cite{Wu-DS}, and are proved to be non-linearizable when acting
on the tau function, which implies that this Drinfeld-Sokolov
hierarchy is not of topological type in the sense of \cite{DZ}. In
the proof we obtain a formula \eqref{OT} to calculate the obstacles
$O_j$. This formula, with its two sides arising from different
contexts, still needs to be better understood. We plan to study it
in follow-up work.

\vskip 0.5truecm \noindent{\bf Acknowledgments.} The authors are
grateful to Youjin Zhang for helpful discussions and comments. They
also thank the referees for their suggestions. The research of
C.-Z.\,W. has received specific funding under the ``Young SISSA
Scientists' Research Projects'' scheme 2012-2013, promoted by the
International School for Advanced Studies (SISSA), Trieste, Italy.


{\small

\begin{thebibliography}{99}\setlength{\itemsep}{-0.2ex}
\bibitem{ASvM} M. Adler, T. Shiota, P. van Moerbeke.
A Lax representation for the vertex operator and the central
extension. Comm. Math. Phys. 171 (1995), no. 3, 547--588.

\bibitem{DJKM-CKP} E. Date,  M. Jimbo,  M. Kashiwara, T. Miwa.
Transformation groups for soliton equations. VI. KP hierarchies of
orthogonal and symplectic type. J. Phys. Soc. Japan 50 (1981), no.
11, 3813--3818.

\bibitem{DKJM-KPBKP}
 E. Date,  M. Kashiwara, M. Jimbo,  T. Miwa. Transformation groups for
soliton equations. Nonlinear integrable systems---classical theory
and quantum theory (Kyoto, 1981), 39--119, World Sci. Publishing,
Singapore, 1983.

\bibitem{Di} L.\,A. Dickey.
On additional symmetries of the KP hierarchy and Sato's B\"{a}cklund
transformation. Comm. Math. Phys. 167 (1995), no. 1, 227--233.

\bibitem{Dickey} L.A. Dickey. Soliton equations and Hamiltonian systems. Second edition.
Advanced Series in Mathematical Physics, 26. World Scientific
Publishing Co., Inc., River Edge, NJ, 2003.

\bibitem{DS} V.\,G. Drinfeld,  V.\,V. Sokolov. Lie algebras and equations of
Korteweg--de Vries type. (Russian) Current problems in mathematics,
Vol. 24, 81--180, Itogi Nauki i Tekhniki, Akad. Nauk SSSR, Vsesoyuz.
Inst. Nauchn. i Tekhn. Inform., Moscow, 1984.

\bibitem{DLZ} B. Dubrovin, S.-Q. Liu,  Y. Zhang. Frobenius manifolds and central invariants for the Drinfeld-Sokolov biHamiltonian structures. (English summary)
Adv. Math. 219 (2008), no.3, 780--837.

\bibitem{DZ}  B. Dubrovin, Y. Zhang.  Normal forms of integrable PDEs, Frobenius manifolds
and Gromov-Witten invariants. Preprint arXiv: math.DG/0108160.

\bibitem{HTFM} J. He, K. Tian, A. Foerster, W.-X. Ma.
Additional symmetries and string equation of the CKP hierarchy.
Lett. Math. Phys. 81 (2007), no. 2, 119--134.

\bibitem{HMSG} T.\,J. Hollowood, J.\,L. Miramontes,  J. S\'anchez
Guill\'en. Additional symmetries of generalized integrable
hierarchies. J. Phys. A 27 (1994), no. 13, 4629--4644.

\bibitem{JM83}  Jimbo, M.; Miwa, T.
Solitons and infinite-dimensional Lie algebras. Publ. Res. Inst.
Math. Sci. 19 (1983), no.3, 943--1001.

\bibitem{Kac} V.\,G. Kac. Infinite-dimensional Lie algebras. Third edition.
Cambridge University Press, Cambridge, 1990, RI, 1989.

\bibitem{KR} V.\,G. Kac, A.\,K. Raina. Bombay lectures on highest weight
representations of infinite-dimensional Lie algebras. Advanced
Series in Mathematical Physics, 2. World Scientific Publishing Co.,
Inc., Teaneck, NJ, 1987. xii+145 pp.

\bibitem{Kon} Kontsevich, M. Intersection theory on the moduli space of curves and the matrix Airy function.
Comm. Math. Phys. 147 (1992), no.1, 1--23.

\bibitem{vdL} J. van de Leur. The Adler--Shiota--van Moerbeke formula for the BKP
hierarchy. J. Math. Phys. 36 (1995), no. 9, 4940--4951.

\bibitem{vdLOS} J. van de Leur, A.\,Yu. Orlov, T. Shiota. CKP Hierarchy, bosonic
tau function and bosonization formulae. SIGMA Symmetry Integrability
Geom. Methods Appl. 8 (2012), Paper 036, 28 pp.

\bibitem{LWZ} S.-Q. Liu, C.-Z. Wu,  Y. Zhang. On the Drinfeld--Sokolov hierarchies of
$D$ type.  Intern. Math. Res. Notices, 2011, no. 8, 1952--1996.

\bibitem{Or}  A.\,Yu. Orlov.
Vertex operator, $\bar\pd$-problem, symmetries, variational
identities and Hamiltonian formalism for 2+1 integrable systems.
Plasma theory and nonlinear and turbulent processes in physics, Vol.
1, 2 (Kiev, 1987), 116--134, World Sci. Publishing, Singapore, 1988.

\bibitem{OS} A.\,Yu. Orlov, E.\,I. Schulman.
Additional symmetries for integrable equations and conformal algebra
representation. Lett. Math. Phys. 12 (1986), no. 3, 171--179.

\bibitem{Tu} M.-H. Tu.
On the BKP hierarchy: additional symmetries, Fay identity and
Adler-Shiota-van Moerbeke formula. Lett. Math. Phys. 81 (2007), no.
2, 93--105.

\bibitem{dV} K. de Vos. Symmetries of integrable hierarchies and matrix model
constraints. Nuclear Phys. B 375 (1992), no.  2, 478--500.

\bibitem{Wi} Witten, E. Two-dimensional gauge theories revisited.
J. Geom. Phys. 9 (1992), no.4, 303--368.

\bibitem{Wu} C.-Z. Wu. From additional symmetries to linearization of Virasoro
symmetries. Physica D, 249 (2013), 25--37.

\bibitem{Wu-DS} C.-Z. Wu. Tau functions and Virasoro symmetries for Drinfeld-Sokolov
    hierarchies. Preprint arXiv: nlin.SI/1203.5750.

\end{thebibliography}

}
\end{document}